\documentclass[aps,prl,reprint,preprintnumbers,twocolumn]{revtex4-2}
\usepackage{graphicx}
\usepackage{bm}
\usepackage{epsfig}
\usepackage{ulem}
\usepackage{color}
\usepackage{mathrsfs}
\usepackage{dcolumn}
\usepackage{setspace}
\usepackage{array}
\usepackage{amsmath}
\usepackage{amssymb}
\usepackage{dsfont}
\usepackage{dcolumn}
\usepackage{multirow}
\usepackage{bibentry,natbib}
\usepackage{booktabs}
\usepackage{appendix}
\usepackage{extarrows}

\begin{document}

\title{Vacuum Torque Without Anisotropy: Switchable Casimir Torque Between Altermagnets}
\author{Zixuan Dai$^1$,  Qing-Dong Jiang$^{1,2}$}
\email{qingdong.jiang@sjtu.edu.cn}
\affiliation{$^1$Tsung-Dao Lee Institute \& School of Physics and Astronomy, Shanghai Jiao Tong University, Pudong, Shanghai, 201210, China\\
$^2$Shanghai Branch, Hefei National Laboratory, Shanghai, 201315, China.
}

\begin{abstract}

Casimir torque is conventionally associated with explicit breaking of rotational symmetry, arising from material dielectric anisotropy, geometric asymmetry, or externally applied fields that themselves break rotational invariance. Here we demonstrate a fundamentally different mechanism: an axially symmetric magnetic field can generate a Casimir torque by inducing an axially asymmetric Casimir energy—and can even reverse the torque’s sign. Focusing on two-dimensional altermagnets, we show that a magnetic field applied perpendicular to the plane—while preserving in-plane rotational symmetry—activates an orientation-dependent vacuum interaction through the combined crystalline symmetry $\rm {C}_n {T}$ inherent to altermagnetic order. The resulting torque emerges continuously and scales quadratically with the magnetic field strength. We further analyze its temperature and distance dependence, revealing scaling behaviors that are qualitatively different from those found in uniaxial bulk materials.
Our results identify time-reversal symmetry breaking as a powerful route for engineering both the sign and strength of Casimir torque and establish altermagnets as an exciting platform for exploring phenomena driven by vacuum quantum fluctuations.
\end{abstract}
\maketitle

\textit{Introduction}: All Casimir phenomena originate from the interplay between vacuum quantum fluctuations and continuous symmetry breaking \cite{1948Casimir, 1956Lifshitz, CasimirPhysics}. In free space, vacuum fluctuations are isotropic and homogeneous; the introduction of material bodies, however, breaks these continuous symmetries, giving rise to a variety of measurable Casimir interactions. For example, breaking translational symmetry produces a distance-dependent Casimir energy, resulting in the familiar Casimir force, while breaking rotational symmetry renders the interaction orientation-dependent, leading to a Casimir torque. Rotational symmetry can be broken either by intrinsic material anisotropy \cite{1978Barash, 2005Munday, 2009LuoJun, 2017Munday, 2018Persson, 2020Chang, 2025Munday} or by geometric asymmetry \cite{2015Parsegian, 2020grating, 2006corrugated, 2009cylinder, 2012cylinder}. Such torques have been observed experimentally \cite{2018MundayExp} and offer promising routes for applications in nanoelectromechanical systems \cite{2019Manjavacas, 2024Shegai, 2022MundayReview}.
Beyond continuous symmetries, fundamental discrete symmetries---particularly parity and time-reversal symmetry---play a key role in determining the magnitude and even the sign of Casimir forces \cite{2006NoGoTheorem, 2022Kruger, 2019QD, 2017Buhmann, 2008Rosa, 2008RosaPRA, Dai_2025}. Their influence is also evident in related fluctuation-induced phenomena, including propulsion forces \cite{jiang2019axial,2021PropulsionForce, 2024Milton, 2025Kruger}, radiative heat transfer \cite{2021RHT, 2022RHT}, quantum atmosphere effects \cite{2019QD-QuantumAtmosphere, ke2023vacuum}, and cavity-modified material properties \cite{2024TBG, 8qx2-xxh2, yang2025emergent, 1tlw-g26r, 2021ChiralCavity}.
Yet, the influence of these fundamental discrete symmetries on the Casimir torque remains largely unexplored.

Altermagnets are a recently discovered class of collinear magnets distinguished by vanishing net magnetization in real space and momentum-dependent spin polarization in reciprocal space, combining features of both ferromagnets and antiferromagnets \cite{2022prxAMreview1, 2022prxAMreview2}. Their anisotropic spin-split electronic bands and ultrafast dynamics have attracted intense interest for spintronic applications \cite{2021SpinSplitter, 2022SpinSplitter1, 2022SpinSplitter2, 2025STT, 2025SpinCurrent, 2025SpinRelax, 2023TMR, 2024TMR, 2025TMRSun, 2025natrev}. From a symmetry perspective, the unique properties of altermagnets arise from the combined $n$-fold crystalline rotational and time-reversal symmetry, $\rm C_nT$. This symmetry underlies a host of unconventional phenomena, including the anomalous Hall effect \cite{2020AHE, 2025AHE}, higher-order nonlinear transport \cite{2024NonlinearTransport, 2025NonlinearTransport}, orientation-sensitive 
$\phi_0$-Josephson junctions \cite{2024JJ, 2024JJSun, 2023JJ}, spin-polarized Andreev reflection \cite{2025AR, 2025AR2, 2023AR}, and Coulomb drag \cite{2025Xie}. Realizing the potential of altermagnets in nanoscale devices, however, requires precise control of the quantum-fluctuation–induced forces between them, as these dispersion interactions fundamentally determine device stability, functionality, and performance.

\begin{figure}[ht]
    \centering
    \includegraphics[width=0.85\linewidth]{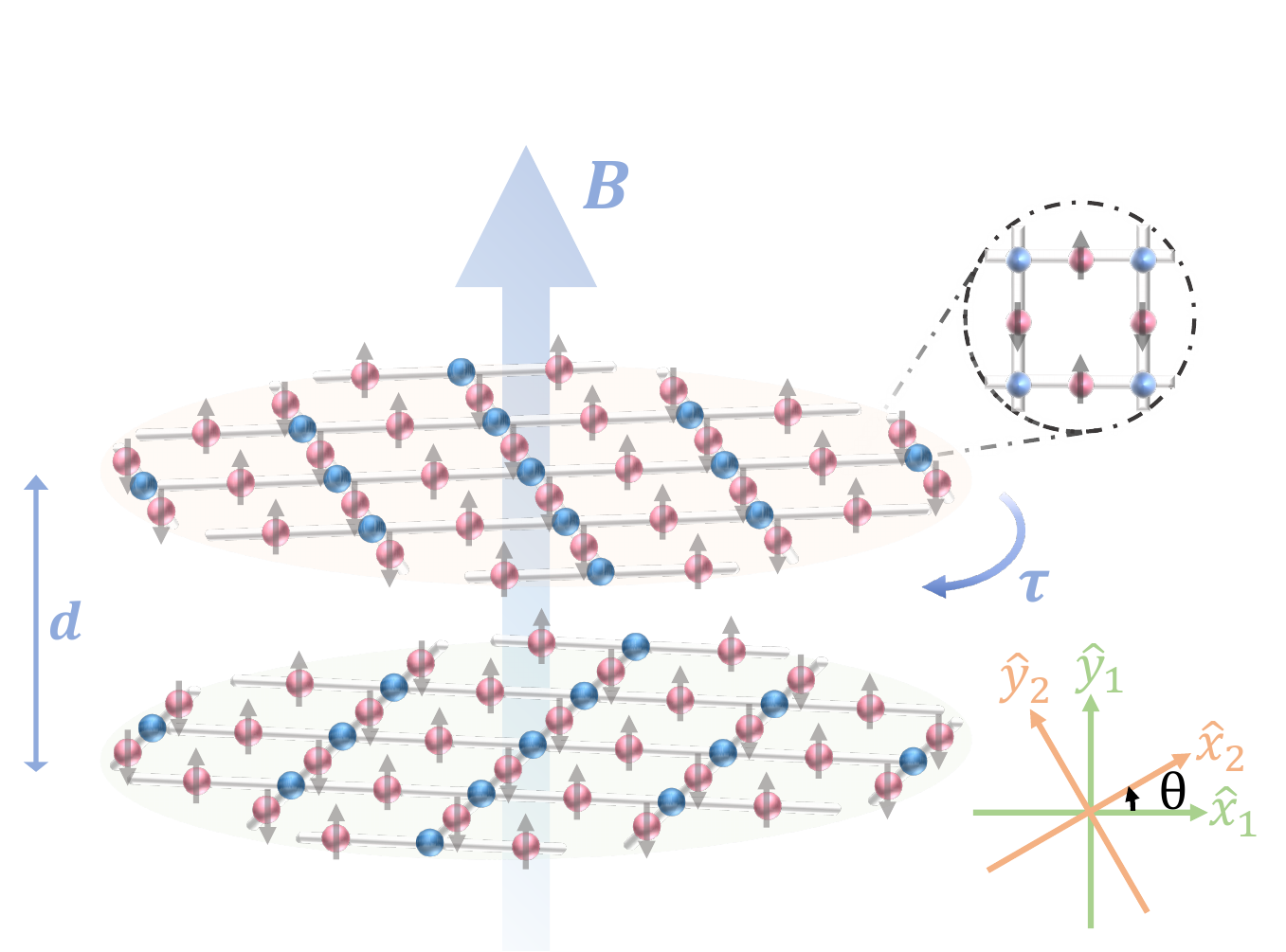}
    \caption{Schematic figure for the Casimir torque between two Lieb-lattice altermagnets with separation distance $d$. The relative angle between the crystal axes is $\theta$. The z-axis is perpendicular to the plane.}
    \label{fig:setup}
\end{figure}

In this work, we demonstrate that time-reversal symmetry breaking offers a novel route to control both the sign and strength of Casimir torque in materials with altermagnetic order (Fig.~1). A magnetic field applied perpendicular to the plane—while preserving in-plane rotational symmetry—activates an orientation-dependent vacuum interaction via the intrinsic $\rm C_nT$ symmetry of altermagnetic materials. The resulting torque emerges continuously from zero and scales quadratically with the field strength. We further explore its temperature and distance dependence, revealing behavior qualitatively distinct from conventional uniaxial bulk materials. 
Together, these results establish that
$\rm C_n T$-symmetric altermagnets provide a versatile platform for engineering tunable Casimir torques via time-reversal–symmetry-breaking fields.

\textit{Altermagnetic Model and Casimir energy}: We consider a system comprises two parallel two-dimensional altermagnets possessing $\rm C_4T$ symmetry, separated by a vacuum gap of distance $d$, as shown in Fig.\ref{fig:setup}. We choose the z-axis to be perpendicular to the plates and align the x- and y-axes with the crystal axes of the first layer. The crystal axes of the second plate is rotated by an angle $\theta$ relative to the crystal axes of the first plate. To incorporate the optical response, we adopt a three-band altermagnetic model for each spin component~\cite{2023_2DAM}, whose Hamiltonian is given by:
\begin{equation}
    H_\sigma(\mathbf{k}) = \left(
    \begin{matrix}
        h_{11,\sigma} & h_{12,\sigma} & h_{13,\sigma} \\
        h^*_{12,\sigma} & h_{22,\sigma} & h_{23,\sigma}\\
        h^*_{13,\sigma} & h^*_{23,\sigma} & h_{33,\sigma}
    \end{matrix}
    \right)
\end{equation}
where
\begin{eqnarray}
    && h_{11,\sigma} = \epsilon_m-\mu-\sigma JS,\; h_{22,\sigma} = \epsilon_{nm}-\mu, \nonumber\\
    && h_{33,\sigma} = \epsilon_m-\mu+\sigma JS, \nonumber\\
    && h_{12,\sigma} = 2t\cos{\frac{k_x}{2}}, \; h_{23,\sigma} = 2t\cos{\frac{k_y}{2}}, \nonumber\\
    && h_{13,\sigma} = 2t_2(\cos{\frac{k_x+k_y}{2}} + \cos{\frac{k_x-k_y}{2}}) \nonumber
\end{eqnarray}
$\epsilon_m$ and $\epsilon_{nm}$ are magnetic and nonmagnetic on-site energy. $t, t_2$ are hopping parameters. $\mu$ is the chemical potential determined by doping. J is the on-site exchange interaction between the electron spin and the localized magnetic moments $S$ at the magnetic site. Since we do not consider the spin-orbit coupling effect, the spin index $\sigma=\pm$ is a good quantum number. 

The Casimir energy between the two altermagnets is given by \cite{2022MundayReview}
\begin{eqnarray}
    E_c(\theta,d) &&= \frac{k_BTA}{4\pi^2}\sum^{\infty}_{n=0}{}' \int_0^\infty k_\parallel dk_\parallel \\
    && \times \int_0^{2\pi} d\phi \; \textup{ln\, det} (\textbf{1}-\textbf{R}_1 \cdot \textbf{R}_2 e^{-2K_nd}) \nonumber
\end{eqnarray}
where $A$ is the plate area and $T$ is the temperature. The summation is performed over the Matsubara frequencies $\xi_n = \frac{2\pi k_BT}{\hbar}$, and the prime denotes that the $n=0$ term is multiplied by a factor of $1/2$. The integration variables $k_\parallel$ and $\phi$ are the radial and angular components of the in-plane wave vector. $K_n = \sqrt{\frac{\xi_n^2}{c^2}+{k^2_\parallel}}$. $\textbf{R}_i$ represents the reflection matrix of plate $i$ ($i = 1,2$), which takes the form
\begin{eqnarray}
\textbf{R}_i &=& \left[
\begin{array}{cc}
r_{i,ss}(i\xi_n, k_\parallel, \phi_i) & r_{i,sp}(i\xi_n, k_\parallel, \phi_i)\\
r_{i,ps}(i\xi_n, k_\parallel, \phi_i) & r_{i,pp}(i\xi_n, k_\parallel, \phi_i)\\
\end{array} \right]
\end{eqnarray}
where $\phi_1 = \phi$ and $\phi_2 = \phi+\theta$. The reflection coefficients $r_{\alpha\beta}(\alpha, \beta\in \{s,p\})$ for a two-dimensional altermagnet are determined by the conductivity tensor $\sigma_{\alpha\beta}$ and explicit expressions are provided in the Supplemental Material ~\cite{SM}. The conductivity tensor is computed within the linear response framework. The Casimir torque is derived by differentiating the Casimir energy with respect to the relative angle $\theta$, which is 
\begin{eqnarray}\label{eq:TorqueGeneral}
    \tau(\theta, d) &=& \frac{k_BTA}{4\pi^2}\sum^{\infty}_{n=0}{}' \int_0^\infty k_\parallel dk_\parallel \int_0^{2\pi} d\phi \; \\
    &\times& \textup{Tr}\left[(1-\textbf{R}_1 \cdot \textbf{R}_2 e^{-2K_nd})^{-1}\textbf{R}_1 \partial_\theta\textbf{R}_2\right]e^{-2K_nd} \nonumber
\end{eqnarray}
Since $\theta$ appears only in the reflection matrix of the second plate, the derivative acts only on $\mathbf{R}_2$.

\begin{figure}[ht]
    \centering
    \includegraphics[width=0.85\linewidth]{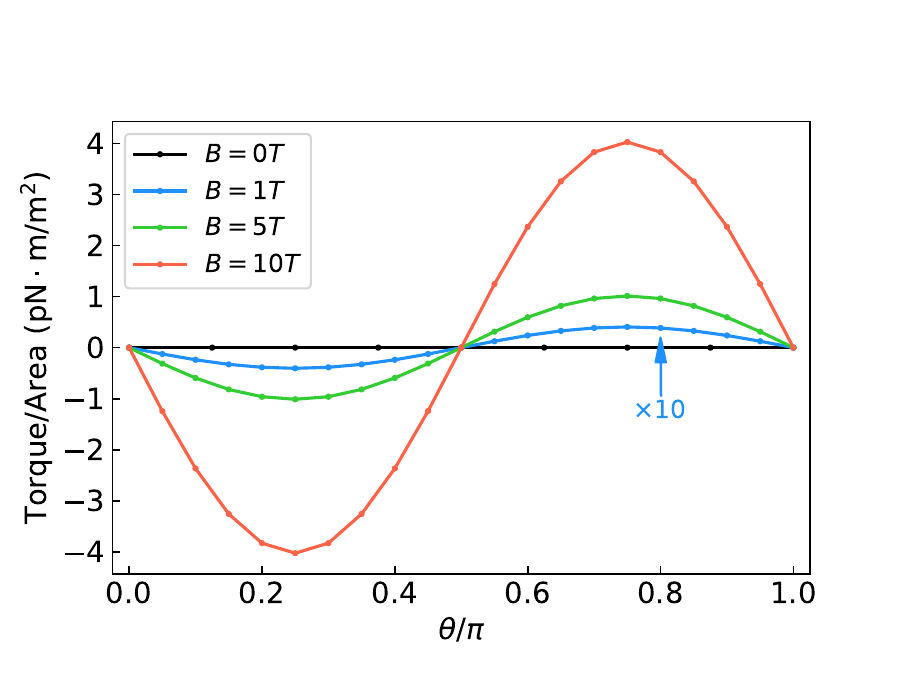}
    \caption{Casimir torque per unit area as a function of $\theta$ between two altermagnets with separation distance $d = 30\textup{nm}$. The temperature is set to be $T = 30\textup{K}$. The torque varies with $\theta$ sinusoidally. The magnitude of the torque increases with the increasing magnetic field.}
    \label{fig:thetadep}
\end{figure}
\textit{Engineering the strength and sign of Casimir torque via a magnetic field}.—
From symmetry considerations, the Casimir torque between two altermagnets vanishes in the absence of external fields. Specifically, the combined $\rm C_4T$ symmetry enforces $\sigma_{xx}(\omega)=\sigma_{yy}(\omega)$ and $\sigma_{xy}(\omega)=-\sigma_{yx}(\omega)$, yielding an isotropic electromagnetic response and thus zero torque.
This situation changes qualitatively upon applying a rotationally invariant magnetic field $B$ along the $z$ axis. Although the field preserves continuous rotational symmetry, it breaks the $\rm C_4T$ symmetry of the altermagnetic state through the Zeeman coupling $\mu_B \sigma B$ in the Hamiltonian. As a result, an orientation-dependent vacuum interaction is activated, allowing a finite Casimir torque to emerge.
We evaluate the optical conductivity and reflection matrices numerically and compute the resulting Casimir torque, shown in Fig.~\ref{fig:thetadep}. As expected, the torque vanishes at $B=0$. With increasing magnetic field strength, a finite torque develops continuously, and its magnitude grows monotonically with $B$, demonstrating a tunable mechanism for controlling both the existence and strength of Casimir torque via time-reversal symmetry breaking. The angular dependence of the torque follows a sinusoidal form in $\theta$, characteristic of weakly anisotropic systems, in agreement with general expectations \cite{2022MundayReview}.

\begin{figure}[t]
    \centering
    \includegraphics[width=0.96\linewidth]{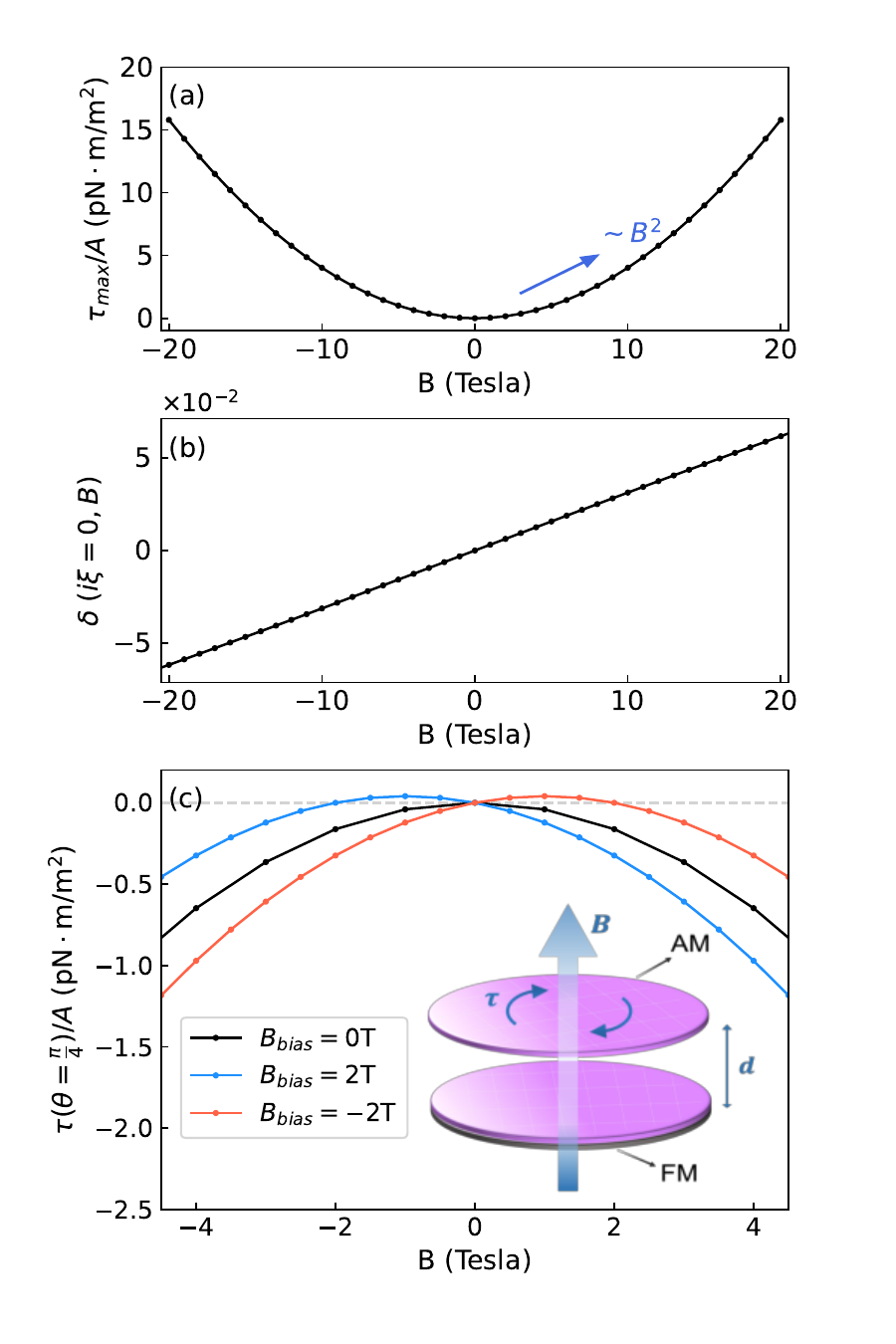}
    \caption{(a) The magnitude of the torque and (b) the degree of anisotropy $\delta = (\sigma_{xx} - \sigma_{yy})/(\sigma_{xx} + \sigma_{yy})$ as a function of the magnetic field $B$ at temperature $T = 30\textup{K}$. The separation is $d = 30\textup{nm}$. When $\mu_BB\ll k_BT$, the torque increases with the magnetic field quadratically, and $\delta$ increases with the magnetic field linearly. Since we are considering two identical altermagnets, the subscript of $\delta$ is omitted here. (c) The torque for $\theta = \pi/4$ at temperature $T = 30\mathrm{K}$ when placing one of the altermagnets (AM) on a ferromagnetic (FM) substrate. The separation between altermagnets is $d = 30\textup{nm}$. The ferromagnetic substrate provide an exchange bias field $B_{bias}$. The torque sign reverses when the external magnetic field $B$ cross two critical points: $B = 0$ and $B = -B_{bias}$.}
    \label{fig:Bdep}
\end{figure}

The magnetic field dependence of the Casimir torque is plotted in Fig.\ref{fig:Bdep} (a), revealing a quadratic scaling $\tau\sim B^2$. The even-in-$B$ behavior originates from the underlying $\rm C_4T$ symmetry, which enforces the relation ${\rm C_4^{-1}} \bar{\bar{\sigma}}(\omega, B) {\rm C_4} = \bar{\bar{\sigma}}(\omega, -B)$. Physically, reversing the magnetic field is equivalent to a $\pi/4$ rotation of the entire system, under which the Casimir interaction remains invariant (see \cite{SM}). 

To quantify the anisotropy of the $i$th altermagnet, it is convenient to define the parameter $\delta_i = (\sigma_{i,xx} - \sigma_{i,yy})/(\sigma_{i,xx} + \sigma_{i,yy})$. The torque is proportional to $\delta_1\delta_2$ at each Matsubara frequency, analogous the case of two biaxially anisotropic slabs \cite{2018Persson}. 
In the weak-field regime, $\mu_B B \ll k_B T$, the anisotropy parameters $\delta_i$ scale linearly with the Zeeman field, as shown in Fig.~\ref{fig:Bdep}(b). This explicit dependence on $\delta_i$ provides a natural and efficient route to control the sign of the torque. For example, if an exchange bias field $B_{\mathrm{bias}}$ is introduced---e.g., by placing one altermagnet on a ferromagnetic substrate \cite{2025ExchangeBias}---the total field acting on that layer becomes $B + B_{\mathrm{bias}}$. The resulting torque then scales as $\tau \propto B(B+B_{\mathrm{bias}})$. Consequently, the Casimir torque reverses sign at $B=0$ and $B=-B_{\mathrm{bias}}$ (see Fig.~\ref{fig:Bdep}(c)). Near zero field, the torque can therefore be switched simply by reversing the direction of the external magnetic field.



\begin{figure}[t]
    \centering
    \includegraphics[width=0.98\linewidth]{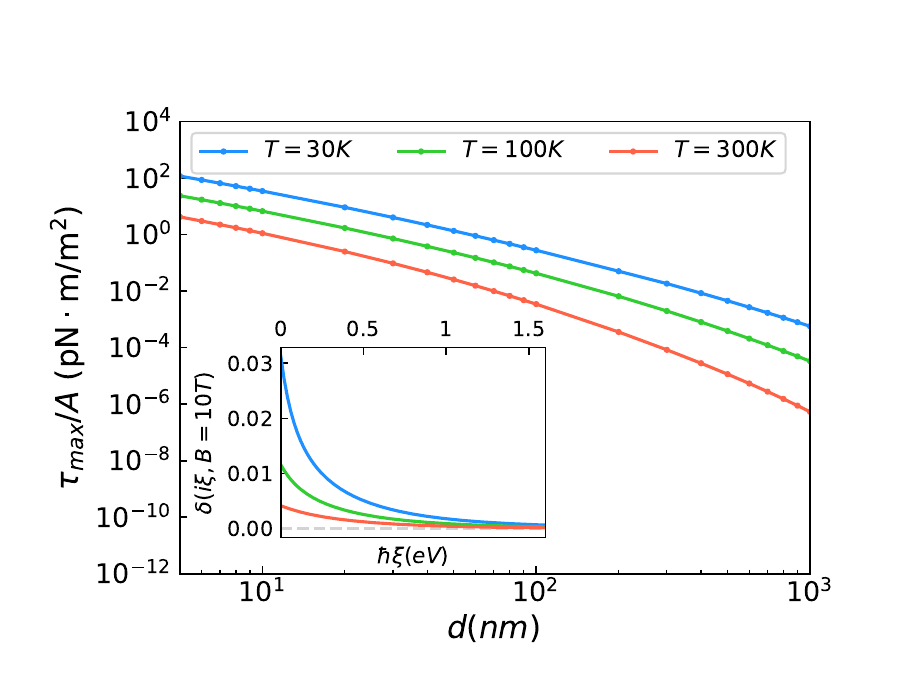}
    \caption{The magnitude of the torque as a function of separation $d$ for temperatures $T = 30\textup{K}$ (blue curve), $100\textup{K}$ (green curve) and $300\textup{K}$ (red curve) in an external magnetic field of $B=10\textup{T}$. At all distances, the torque decreases with increasing temperature. Inset figure depicts the degree of anisotropy $\delta$ as a function of the imaginary frequency at $B = 10T$. $\delta$ decreases monotonically with increasing frequency. For all frequencies, $\delta$ decreases with increasing temperature.}
    \label{fig:T_and_d}
\end{figure}

\textit{Thermal effects and distance scaling}:
We now investigate the influence of temperature on the Casimir torque and its dependence on the separation distance. Fig.~\ref{fig:T_and_d} presents the Casimir torque as a function of the separation $d$ at temperatures $T = 30 \mathrm{K}$, $100 \mathrm{K}$, and $300 \mathrm{K}$ under an external magnetic field of $B = 10 \mathrm{T}$. Over the entire distance range considered, the torque decreases monotonically with increasing temperature. This reduction primarily originates from the suppression of optical anisotropy at elevated temperatures, as illustrated in the inset of Fig.~\ref{fig:T_and_d}.

Notably, this behavior is fundamentally different from that reported in Ref.~\cite{2025Munday}. There, the dielectric functions of the materials are assumed to be temperature independent, and thermal effects arise purely from electromagnetic field fluctuations, becoming relevant only at separations beyond the thermal wavelength $\hbar c/(2\pi k_B T)\approx1 \mu\mathrm{m}$. In contrast, our model demonstrates that temperature directly modifies the optical anisotropy of altermagnets, leading to a substantial modification of the Casimir torque even at submicron distances.

For all temperatures, the magnitude of the torque monotonically decreases with increasing distance. In the following, we consider two identical altermagnets, $\delta_1\delta_2 = \delta_1^2>0$ for all frequencies and there is no sign reversal of the torque as the separation distance varies \cite{2018Persson}. For weakly anisotropic uniaxial bulk materials, the Casimir torque scales as $1/d^2$ in both the non-retarded limit and the high temperature limit, whereas it exhibits a dependence $1/d^3$ in the retarded limit \cite{1978Barash, 2017Munday, 2025Munday}. However, the distance scaling behavior for weakly anisotropic two-dimensional materials, as in our case, is distinguishably different. 
While the retarded and low temperature regime recovers a $1/d^3$ scaling, the torque in the non-retarded limit deviates fundamentally, exhibiting no simple power law, and decays as $\exp(-2\xi_1 d / c)/d^3$ in  high temperature limit. 

To understand these scaling behaviors, it is useful to compare three characteristic energy scales in the system: the thermal energy $k_B T$, the photon energy associated with separation $\hbar c / d$, and the cutoff frequency of the material response $\hbar \omega_0$. In our case, the optical response $\hbar \omega_0 \sim 1~\mathrm{eV}$ far exceeds the thermal energy $k_B T \approx 0.026~\mathrm{eV}$ at $T = 300~\mathrm{K}$. Consequently, depending on the separation, only three limiting scenarios need to be considered, as detailed below.

(i) In the non-retarded limit, i.e. $\hbar{c}/{d}\gg\hbar\omega_0\gg k_BT$, the torque is mainly contributed by the TM wave and the approximated formula can be written as (see \cite{SM})
\begin{eqnarray}\label{eq:non-retarded}
    \tau(\theta, d) = && - \frac{k_BTA}{2\pi d^2}\sum^{\infty}_{n=0}{}' \int_0^\infty \tilde{K}d\tilde{K}\\
    && \times \frac{h_{1,pp}h_{2,pp}e^{-2\tilde{K}}}{(1-g_{1,pp}g_{2,pp}e^{-2\tilde{K}})^2}\delta_1\delta_2\sin{2\theta} \nonumber
\end{eqnarray}
with
\begin{eqnarray}
    && g_{i,pp} = \frac{\tilde{K}\tilde{\sigma}_{t,i}}{\tilde{K}\tilde{\sigma}_{t,i} + \frac{4\xi_nd}{c}}\\
    && h_{i,pp} = \frac{\tilde{K}\tilde{\sigma}_{t,i}\cdot\frac{4\xi_nd}{c}}{(\tilde{K}\tilde{\sigma}_{t,i} + \frac{4\xi_nd}{c})^2}
\end{eqnarray}
where $\tilde{K} = Kd$ and $\tilde{\sigma}_{t,i} = (\sigma_{xx,i} + \sigma_{yy,i}) / \sigma_0$. $\sigma_0 = \sqrt{\epsilon_0 / \mu_0}$ is the vacuum admittance. In the non-retarded limit, the dimensionless variable $\xi_n d/c\ll1$. Unlike uniaxial bulk materials \cite{2017Munday}, the analysis here is complicated by the fact that $\tilde{\sigma}_t$ is also a small parameter, which prevents the naive omission of the term $\xi_n d/c$ in the denominator. Consequently, Eq. (\ref{eq:non-retarded}) can not be expressed as a Laurent polynomial function of $\tilde{K}$ and integration can not be calculated analytically as in Ref.\cite{2017Munday}. This makes the torque scaling ambiguous. Moreover, the scaling of the torque in the non-retarded regime remains nearly identical across different temperatures. This is because temperature alters the interaction primarily by modifying the optical anisotropy $\delta_i$, which exhibits no distance dependence.

(ii) In the retarded, high-temperature limit, i.e. $\hbar\omega_0 \gg k_BT\gg\hbar {c}/{d} $, the dominant contribution to the torque comes from the $n=1$ Matsubara term in Eq. (\ref{eq:TorqueGeneral}). Differing from the case of bulk anisotropic materials, the reflection coefficients for two-dimensional altermagnets at the lowest Matsubara term $n = 0$ are $r_{ss}=r_{sp}=r_{ps}=0$, $r_{pp}=1$, having no angular dependence. Therefore, the $n = 0$ term yields a vanishing contribution to the torque. Taking into account only the $n=1$ Matsubara term and using the one-reflection approximation, the formula for the torque is given by (see \cite{SM}) 
\begin{eqnarray}
\tau_c(\theta) \simeq && -\frac{\hbar c A}{64 \pi^2} \frac{e^{-2\frac{\xi_1}{c}d}}{d^3} \nonumber \\
&&\times \frac{(\sigma_{xx,1} - \sigma_{yy,1})(\sigma_{xx,2} - \sigma_{yy,2})}{\sigma_0^2}\sin{2\theta}
\end{eqnarray}
where $\sigma_0 = \sqrt{\epsilon_0 / \mu_0}$ and $\sigma_{ij}=\sigma_{ij}(i\xi_1)$.
This simple expression clearly reveals that in the high temperature regime, the Casimir torque follows a characteristic scaling of $\exp(-2\xi_1 d / c)/d^3$, and the magnitude of the torque depends linearly on the optical anisotropy $(\sigma_{i,xx}(i\xi_1) - \sigma_{i,yy}(i\xi_1))$ of each altermagnet evaluated at the first Matsubara frequency.

(iii) In the retarded, low-temperature limit, $\hbar\omega_0\gg\hbar{c}/{d}\gg k_B T$, 
the summation over the Matsubara frequency in Eq.(\ref{eq:TorqueGeneral}) can be replaced by the integration over continuous variable $\xi$.
Within the range $200 {\rm nm} \ll d \ll 80  {\rm \mu m}$, the torque follows a power law scaling as $1/d^3$.  This scaling behavior can be understood by introducing dimensionless variables $\tilde{\xi} = \frac{\xi}{c}d, \tilde{k}_\parallel = k_\parallel d$, which remove the distance dependence from the exponential term. After this variable substitution, the conductivity becomes $\sigma(i\frac{\tilde{\xi}}{c}d)\simeq\sigma(0)$ and the reflection matrix $\textbf{R}_i(i\frac{\tilde{\xi}}{d}c, \frac{\tilde{k}_\parallel}{d}, \phi_i) = \textbf{R}_i(i\tilde{\xi}c, \tilde{k}_\parallel, \phi_i)$, both of which are independent of separation $d$. As a result, the
integrand shows no distance dependence anymore, leading to a characteristic $1/d^3$ scaling of the Casimir torque in the retarded regime.


\textit{Conclusions}: In this work, we identify materials with combined crystalline and time-reversal symmetries as a versatile platform for engineering Casimir torque. Using two-dimensional altermagnets as a concrete example, we show that a magnetic field applied perpendicular to the plane—while preserving in-plane rotational symmetry—activates an orientation-dependent vacuum interaction through the intrinsic $\rm C_n T$ symmetry. The resulting torque emerges continuously, scales quadratically with the field strength, and exhibits temperature- and distance-dependent behaviors that differ qualitatively from those of conventional uniaxial bulk materials. Our results establish 
$\rm C_n T$-symmetric systems as an ideal setting for controllably tuning both the strength and the sign of Casimir torque via time-reversal–symmetry breaking.
Our findings offer a practical and minimally invasive route to realizing switchable vacuum-induced torques in two-dimensional quantum materials, facilitating their integration into tunable nanoscale rotational elements and torque-sensitive nanomechanical devices.

\begin{acknowledgments}
\textit{Acknowledgments}: The work is supported by National Natural Science Foundation of China (NSFC) under Grant No.12374332 and Innovation Program for Quantum Science and Technology Grant No.2021ZD0301900, and Shanghai Science and Technology Innovation Action Plan Grant No. 24LZ1400800.
\end{acknowledgments}

\bibliography{ref} 

\end{document}


\title{Supplemental Materials for ``Vacuum Torque Without Anisotropy: Switchable Casimir Torque Between Altermagnets''}
\author{Zixuan Dai$^1$,  Qing-Dong Jiang$^{1,2}$}
\affiliation{$^1$Tsung-Dao Lee Institute \& School of Physics and Astronomy, Shanghai Jiao Tong University, Pudong, Shanghai, 201210, China\\
$^2$Shanghai Branch, Hefei National Laboratory, Shanghai, 201315, China.
}

\maketitle

\paragraph*{Outline of the Supplemental Material.}
The Supplemental Material is organized into three sections.
The first section presents the symmetry analysis, band structure, and optical conductivity of two-dimensional altermagnets.
The second section provides detailed calculations of the Casimir torque, including its asymptotic behavior in different distance regimes.
The third section contains a detailed derivation of the reflection coefficients for an altermagnetic thin film.

\section{I. Symmetry, band structure and conductivity of 2D Altermagnet}
In this section, we introduce the Hamiltonian for a 2D altermagnet with $C_4T$ symmetry, and show the band structure and conductivity.

To investigate the Casimir torque between altermagnets induced by time-reversal symmetry breaking, we employ the Lieb-lattice model for 2D altermagnets with $C_4T$ symmetry as in Ref.\cite{2023_2DAM}. The tight-binding Hamiltonian 
\begin{equation}
    H_\sigma(\mathbf{k}) = \left(
    \begin{matrix}
        \epsilon_m-\mu-\sigma JS & 2t\cos{\frac{k_x}{2}} & 2t_2(\cos{\frac{k_x+k_y}{2}} + \cos{\frac{k_x-k_y}{2}}) \\
        2t\cos{\frac{k_x}{2}} & \epsilon_{nm}-\mu & 2t\cos{\frac{k_y}{2}} \\
        2t_2(\cos{\frac{k_x+k_y}{2}} + \cos{\frac{k_x-k_y}{2}}) & 2t\cos{\frac{k_y}{2}} & \epsilon_m-\mu+\sigma JS
    \end{matrix}
    \right)
\end{equation}
$\epsilon_m$ and $\epsilon_{nm}$ are magnetic and nonmagnetic on-site energy. $t, t_2$ are hopping parameters. $\mu$ is the chemical potential determined by doping. Since we do not consider the spin-orbit coupling effect, the spin index $\sigma=\pm$ is a good quantum number. J is the on-site exchange interaction between the electron spin and the localized magnetic moments $S$ at the magnetic site. 

\begin{figure}[ht]
    \centering
    \includegraphics[width=0.5\linewidth]{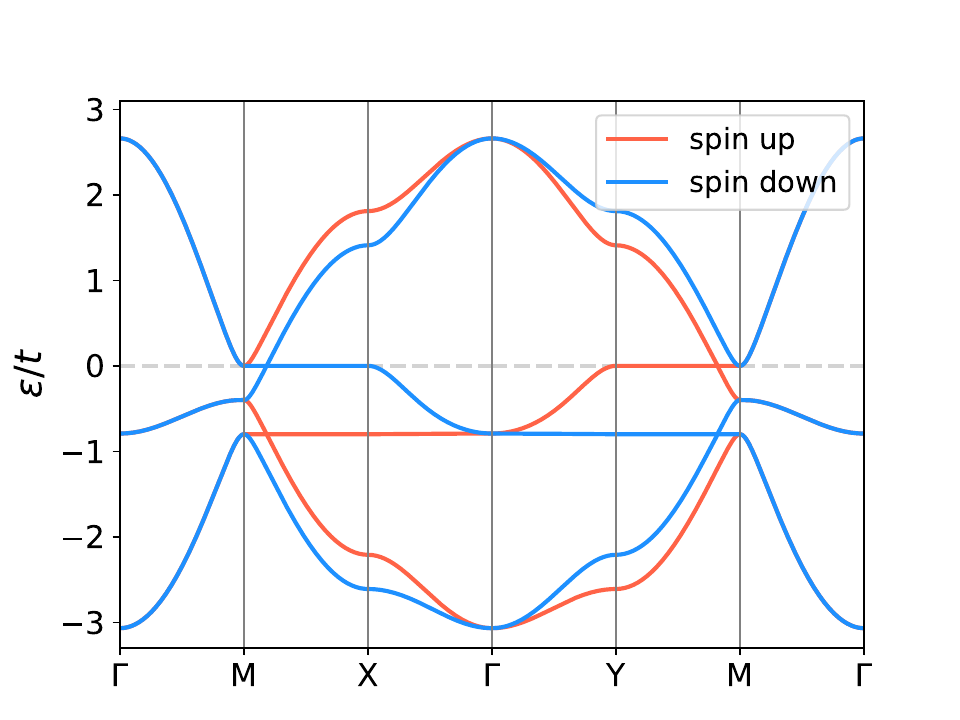}
    \caption{The energy band structure of the altermagnet. Red and blue curves correspond to the energy dispersions of spin-up and spin-down electrons, respectively. The parameters used are: $t = 1, t_2 = 0.1, E_m = 0.0, E_{nm} = 0.0, \mu = 0.4, JS = 0.4$. All values are in eV, but units are omitted for clarity.}
    \label{fig_band}
\end{figure}

The band structure of this 2D altermagnet is calculated and shown in Fig.\ref{fig_band}. Breaking the $PT$ symmetry, the energy bands present a characteristic large spin splitting. On the other hand, the operations in symmetry group $\{I, C_{2z}, M_x, M_y, C^+_{4z}T$, $C^-_{4z}T, M_{xy}T, M_{x\bar{y}}T\}$ guarantee the relations:
\begin{eqnarray}
    C_{2z} : E_\sigma(k_x, k_y) &=& E_\sigma(-k_x, -k_y)\nonumber\\
    M_x : E_\sigma(k_x, k_y) &=& E_\sigma(k_x, -k_y)\nonumber\\
    M_y : E_\sigma(k_x, k_y) &=& E_\sigma(-k_x, k_y)\nonumber\\
    C^+_{4z}T : E_\sigma(k_x, k_y) &=& E_{-\sigma}(k_y, -k_x)\\
    C^-_{4z}T : E_\sigma(k_x, k_y) &=& E_{-\sigma}(-k_y, k_x)\nonumber\\
    M_{xy}T : E_\sigma(k_x, k_y) &=& E_{-\sigma}(-k_x, -k_y)\nonumber\\
    M_{x\bar{y}}T : E_\sigma(k_x, k_y) &=& E_{-\sigma}(k_x, k_y)\nonumber
\end{eqnarray}
Here $T$ is the time-reversal operator. $C_{2z}, C^+_{4z}T, C^-_{4z}T$ are $\pi$ rotation, $\pi/2$ counterclockwise rotation, and $\pi/2$ clockwise rotation. $M_x, M_y,M_{xy}, M_{x\bar{y}}$ are mirror operations with respect to the x-axis, the y-axis, and the diagonals. Due to the $M_{xy}$ and $M_{x\bar{y}}$ symmetries, the energy bands remain spin-degenerate along the $\Gamma$-$M$ direction. 

\begin{figure}[h]
    \centering
    \includegraphics[width=0.75\linewidth]{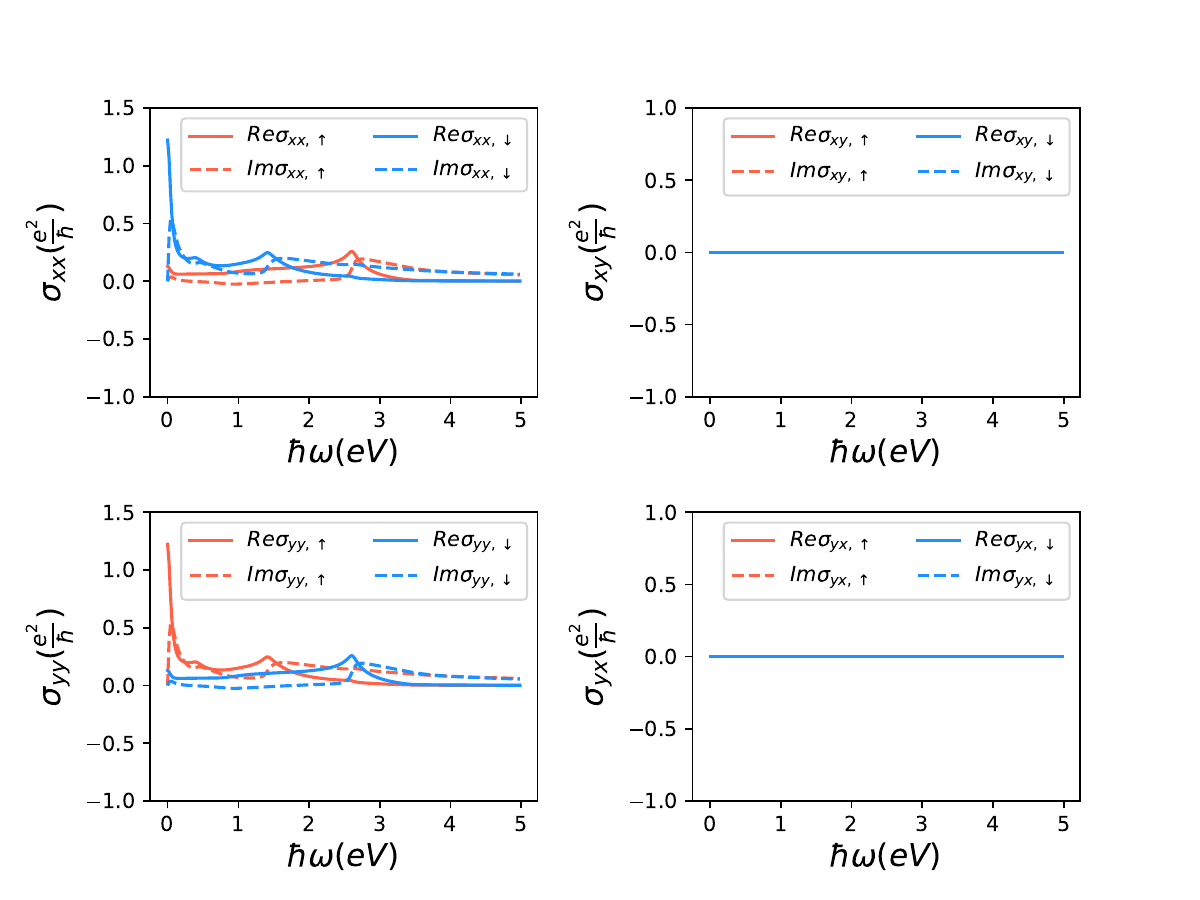}
    \caption{Conductivity of the altermagnet at real frequency at temperature $T = 300K$ without magnetic field. Red and blue curves denote the conductivity for spin-up and spin-down electrons, respectively, while solid and dashed lines represent the real and imaginary parts.  The right column shows subfigures with vanishing non-diagonal conductivity. Parameters are consistent with Fig.\ref{fig_band}, using $\gamma = 0.05$eV.}
    \label{fig_conductivity}
\end{figure}

\begin{figure}[h]
    \centering
    \includegraphics[width=0.5\linewidth]{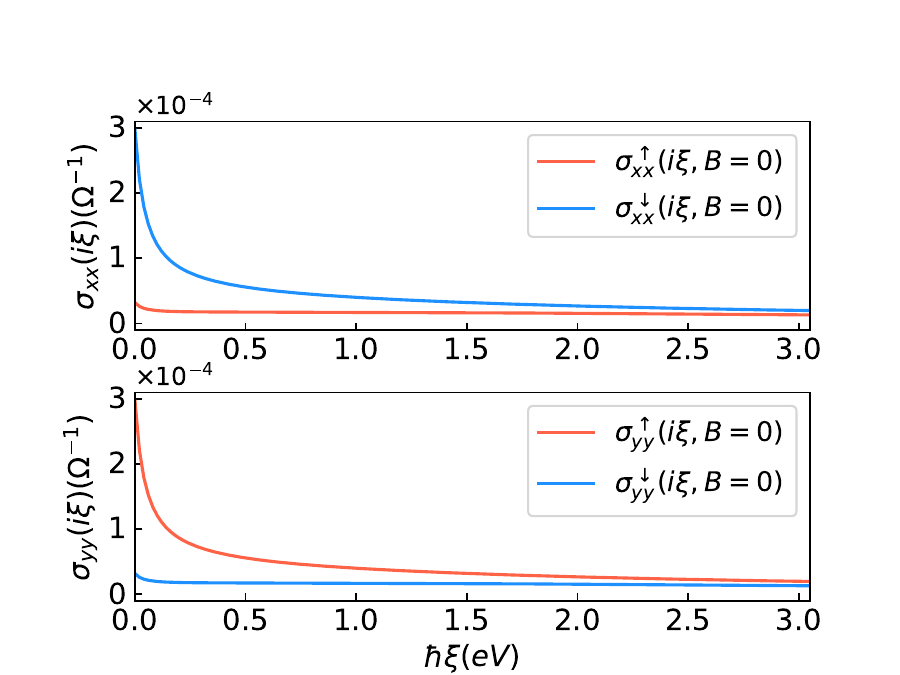}
    \caption{Conductivity of the altermagnet at imaginary frequency at temperature $T = 300K$ without magnetic field. The non-diagonal conductivities vanish. Parameters are consistent with Fig.\ref{fig_band}, using $\gamma = 0.05$eV.}
    \label{fig:conductivity_imagfreq}
\end{figure}

The conductivity $\sigma_{\alpha\beta}(\alpha,\beta \in\{x, y\})$ of the altermagnet is calculated numerically by applying the Kubo formula:
\begin{equation}\label{Kubo}
    \sigma_{\alpha\beta}(\omega) = \frac{-i}{N\Omega}\frac{e^2}{\hbar}\sum_{\mathbf{k}mn}\frac{f_{m\mathbf{k}}-f_{n\mathbf{k}}}{E_m(\mathbf{k})-E_n(\mathbf{k})} \frac{\bra{\mathbf{k}m}\frac{\partial H}{\partial k_\alpha}\ket{\mathbf{k}n}\bra{\mathbf{k}n}\frac{\partial H}{\partial k_\beta}\ket{\mathbf{k}m}}{\hbar\omega + E_m(\mathbf{k}) - E_n(\mathbf{k}) + i\gamma}
\end{equation}
where $m, n\in\{1,2,3\}$ are band index and $\gamma$ is the damping factor. $f_{m\mathbf{k}} = 1/(e^{{E_m(\mathbf{k})}/{k_BT}}+1)$ is the Fermi-Dirac distribution function at energy $E_m(\mathbf{k})$. $N$ is the number of unit cells in the crystal, and $\Omega$ is the area of each unit cell. When $m = n$, the first term will be evaluated by the derivative of the Fermi-Dirac distribution function $df(E)/dE$.
Fig.\ref{fig_conductivity} and Fig.\ref{fig:conductivity_imagfreq} show the conductivities of altermagnets at real and imaginary frequency at $T = 300K$ without magnetic field. The $C_4^+T$ (or $C_4^-T$) symmetry guarantees $\sigma^\uparrow_{xx}(\omega) = \sigma^\downarrow_{yy}(\omega), \sigma^\downarrow_{xx}(\omega) = \sigma^\uparrow_{yy}(\omega)$. The $M_x$ (or $M_y$) symmetry guarantees $\sigma^\uparrow_{xy}(\omega) = \sigma^\downarrow_{xy}(\omega) = \sigma^\uparrow_{yx}(\omega) = \sigma^\downarrow_{yx}(\omega) = 0$. Therefore, the total conductivity $\sigma_{xx}(\omega) = \sigma_{yy}(\omega), \sigma_{xy}(\omega) = \sigma_{yx}(\omega) = 0$. Without an applied magnetic field, the altermagnet exhibits an isotropic conductivity tensor. Consequently, no Casimir torque arises since the Casimir energy between two such altermagnets shows no angular dependence. To induce the Casimir torque, we can break the $C_4T$ symmetry through the application of an external magnetic field, an in-plane electric field, or the introduction of spin-orbit effects.

\section{II. Casimir Torque in different distance Regimes}
This section presents a general formula for the Casimir torque between two-dimensional altermagnets based on their reflection coefficients. We then derive its expression in the weak-anisotropy regime and further obtain simplified forms in the non-retarded, retarded, and high-temperature limits. Investigating the Casimir torque in each limit helps clarify the characteristic distance scaling of the torque across different separation regimes.

\subsection{A. General formula}
The Casimir energy of the system at temperature $T$ is given by \cite{2022MundayReview}
\begin{equation}\label{eq:CasimirEnergy}
E_c(\theta,d)= \frac{k_BTA}{4\pi^2}\sum^{\infty}_{n=0}{}' \int_0^\infty k_\parallel dk_\parallel \int_0^{2\pi} d\phi \; \textup{ln\, det} (1-\textbf{R}_1 \cdot \textbf{R}_2 e^{-2K_nd})
\end{equation}
where $A$ is the plate area, $d$ is the separation distance and $\theta$ represents the relative angle between the crystal axes of the AM. The summation is performed on the Matsubara frequencies $\xi_n = \frac{2\pi k_BT}{\hbar}$, and the prime denotes that the $n=0$ term is multiplied by a factor of $1/2$. The integration variables $k_\parallel$ and $\phi$ are the radial and angular components of the in-plane wave vector. $K_n = \sqrt{\frac{\xi_n^2}{c^2}+{k^2_\parallel}}$. $\textbf{R}_i$ represents the reflection matrix of plate $i$ ($i = 1,2$), which takes the form
\begin{eqnarray}
\textbf{R}_i &=& \left[
\begin{array}{cc}
r_{i,ss}(i\xi_n, k_\parallel, \phi_i) & r_{i,sp}(i\xi_n, k_\parallel, \phi_i)\\
r_{i,ps}(i\xi_n, k_\parallel, \phi_i) & r_{i,pp}(i\xi_n, k_\parallel, \phi_i)\\
\end{array} \right]
\end{eqnarray}
where $\phi_1 = \phi$ and $\phi_2 = \phi+\theta$. 

Taking the derivative with respect to $\theta$, we obtain the formula for the Casimir torque
\begin{eqnarray}\label{eq:TorqueGeneral}
    \tau(\theta, d) &=& - \frac{\partial E(\theta, d)}{\partial\theta} \\
    &=& \frac{k_BTA}{4\pi^2}\sum^{\infty}_{n=0}{}' \int_0^\infty k_\parallel dk_\parallel \int_0^{2\pi} d\phi \; 
    \textup{Tr}\left[(1-\textbf{R}_1 \cdot \textbf{R}_2 e^{-2K_nd})^{-1}\textbf{R}_1 \partial_\theta\textbf{R}_2\right]e^{-2K_nd}
\end{eqnarray}
We use the property that $\frac{d}{dt}|M(t)| = \frac{1}{|M|}\mathrm{Tr}\left[M^{-1}\frac{dM}{dt}\right]$, where $M(t)$ is a matrix depending on the parameter $t$ and $|M(t)|$ is its determinant. Since $\theta$ appears only in the reflection matrix of the second plate, the derivative acts only on $\mathbf{R}_2$. Or equivalently, the Casimir torque
\begin{eqnarray}\label{eq:torque}
\tau_c(\theta,d) = && \frac{k_BTA}{4\pi^2}\sum^{\infty}_{n=0}{}'
\int_0^\infty k_\parallel dk_\parallel \int_0^{2\pi} d\phi \; \\
\times &&
\frac{(r^1_{ss}\dot{r}^2_{ss}+r^1_{sp}\dot{r}^2_{ps}+r^1_{ps}\dot{r}^2_{sp}+r^1_{pp}\dot{r}^2_{pp})e^{-2Kd} - (r^1_{ss}r^1_{pp} - r^1_{sp}r^1_{ps})(r^2_{ss}\dot{r}^2_{pp} + \dot{r}^2_{ss}r^2_{pp}-r^2_{sp}\dot{r}^2_{ps}-\dot{r}^2_{sp}r^2_{ps})e^{-4Kd}}
{1-(r^1_{ss}r^2_{ss}+r^1_{sp}r^2_{ps}+r^1_{ps}r^2_{sp}+r^1_{pp}r^2_{pp})e^{-2Kd}+(r^1_{ss}r^1_{pp}-r^1_{sp}r^1_{ps})(r^2_{ss}r^2_{pp}-r^2_{sp}r^2_{ps})e^{-4Kd}}\nonumber
\end{eqnarray}
where $\dot{r}_{\alpha\beta} \equiv \frac{\partial r_{\alpha\beta}}{\partial \theta}(\alpha, \beta\in \{s,p\})$ is the derivative of reflection coefficients with respect to angle $\theta$.

The general reflection coefficients of two-dimensional materials are derived in the next section and are given by
\begin{eqnarray}\label{eq:Rcoeff}
    &&{r}_{i,ss}(i\xi_n, k_\parallel, \phi_i) = \frac{1}{\Delta_{i}}\left[-\frac{\xi_n^2}{c^2}\tilde{\sigma}_{t,i}(1 - \delta_i\cos{2\phi_i}) - \frac{\xi_n K_n}{4c}\tilde{\sigma}_{t,i}^2(1-\delta_i^2)\right] \nonumber \\  
    &&{r}_{i,pp}(i\xi_n, k_\parallel, \phi_i) = \frac{1}{\Delta_{i}}\left[\frac{\xi_n K_n}{4c}\tilde{\sigma}_{t,i}^2(1-\delta_i^2) + K_n^2\tilde{\sigma}_{t,i}(1 + \delta_i\cos{2\phi_i})\right] \nonumber \\
    &&{r}_{i,ps}(i\xi_n, k_\parallel, \phi_i) = \frac{1}{\Delta_{i}}\left[-\frac{\xi_n K_n}{c}\tilde{\sigma}_{t,i}\delta_i \sin{2\phi_i}\right] \\
    &&{r}_{i,sp}(i\xi_n, k_\parallel, \phi_i) = \frac{1}{\Delta_{i}}\left[\frac{\xi_n K_n}{c}\tilde{\sigma}_{t,i}\delta_i \sin{2\phi_i}\right] \nonumber \\
\end{eqnarray}
with
\begin{eqnarray}
    \Delta_{i}(i\xi_n, k_\parallel, \phi_i) = \frac{\xi_n^2}{c^2}\tilde{\sigma}_{t,i}(1 - \delta_i\cos{2\phi_i}) + \frac{\xi_n K_n}{4c}\tilde{\sigma}_{t,i}^2(1-\delta_i^2) + K_n^2\tilde{\sigma}_{t,i}(1 + \delta_i\cos{2\phi_i}) + \frac{4\xi_n K_n}{c}
\end{eqnarray}
where we define the degree of anisotropy $\delta_i(i\xi_n) = \frac{\sigma_{xx,i} - \sigma_{yy,i}}{\sigma_{xx,i} + \sigma_{yy,i}}$ and the normalized total diagonal conductivity $\tilde{\sigma}_{t,i}(i\xi_n) = (\sigma_{xx,i} + \sigma_{yy,i})/\sigma_0$ with $\sigma_0 = \sqrt{\epsilon_0/\mu_0}$. The information of conductivities $\sigma_{xx,i}$ and $\sigma_{yy,i}$ is now incorporated in the parameters $\delta_i$ and $\tilde{\sigma}_{t,i}$.

\subsection{B. Weak anisotropy limit}
In the weak anisotropy regime, i.e. $\delta_i\ll1$, we can expand the Casimir energy (Eq.(\ref{eq:CasimirEnergy})) to the second order in $\delta_i$
\begin{equation}\label{eq:CasimirEnergy_2ndOrder}
E_c(\theta,d) = \frac{k_BTA}{4\pi^2}\sum^{\infty}_{n=0}{}' \int_0^\infty k_\parallel dk_\parallel \int_0^{2\pi} d\phi \; \left[\ln \mathcal{D}^{(0)} + \frac{\mathcal{D}^{(1)}}{\mathcal{D}^{(0)}} + \frac{\mathcal{D}^{(2)}}{\mathcal{D}^{(0)}} - \frac{1}{2}\left(\frac{\mathcal{D}^{(1)}}{\mathcal{D}^{(0)}}\right)^2\right] + \mathcal{O}(\delta_i^3)
\end{equation}
where $\mathcal{D} = \textbf{1}-\textbf{R}_1 \cdot \textbf{R}_2 e^{-2K_nd}$ and $\mathcal{D}^{(i)}$ is the $i$th-order part of $\mathcal{D}$. Explicitly, $\mathcal{D}$ is the denominator in Eq.(\ref{eq:torque}).

In reflection coefficients, terms that depend on the angle are always accompanied by $\delta_i$ (See Eq.(\ref{eq:Rcoeff})). On the other hand, to obtain a non-vanishing $\theta$-dependent contribution at second order after integrating over $\phi$, cross terms such as $\cos(2\phi)\cos(2\phi + 2\theta)$ or $\sin(2\phi)\sin(2\phi + 2\theta)$ are required. Therefore, it is sufficient to expand the reflection coefficients of each altermagnet to the first order in $\delta_i$. Finally, the torque at each Matsubara frequency will be proportional to $\delta_1\delta_2$ and exhibit the $\sin(2\theta)$ angular dependence, which is typical for systems with weak anisotropy. 

Expanding the reflection coefficients to the first order of $\delta_i$, we have 
\begin{eqnarray}\label{eq:RcoeffSmall}
    r_{i,ss}(i\xi_n, k_\parallel, \phi) = -\frac{1}{\tilde{\Delta}_{i,\delta}}\left[\frac{\xi_n^2}{c^2}\tilde{\sigma}_{t,i}(1 - \delta_i\cos{2\phi_i}) + \frac{\xi_n K_n}{4c}\tilde{\sigma}_{t,i}^2\right] 
    &&+ \frac{1}{\tilde{\Delta}^2_{i,\delta}}\left[\frac{\xi_n^2}{c^2}\tilde{\sigma}_{t,i} + \frac{\xi_n K_n}{4c}\tilde{\sigma}_{t,i}^2\right]\left[-\frac{\xi_n^2}{c^2}\tilde{\sigma}_{t,i} + K_n^2\tilde{\sigma}_{t,i}\right]\delta_i\cos{2\phi_i} \nonumber\\
    r_{i,pp}(i\xi_n, k_\parallel, \phi) = \frac{1}{\tilde{\Delta}_{i,\delta}}\left[\frac{\xi_n K_n}{4c}\tilde{\sigma}_{t,i}^2 + K_n^2\tilde{\sigma}_{t,i}(1 + \delta_i\cos{2\phi_i})\right] 
    &&- \frac{1}{\tilde{\Delta}^2_{i,\delta}}\left[\frac{\xi_n K_n}{4c}\tilde{\sigma}_{t,i}^2 + K_n^2\tilde{\sigma}_{t,i}\right]\left[-\frac{\xi_n^2}{c^2}\tilde{\sigma}_{t,i} + K_n^2\tilde{\sigma}_{t,i}\right]\delta_i\cos{2\phi_i} \nonumber\\
    r_{i,ps}(i\xi_n, k_\parallel, \phi) = -\frac{1}{\tilde{\Delta}_{i,\delta}}&&\left[\frac{\xi_n K_n}{c}\tilde{\sigma}_{t,i}\delta_i \sin{2\phi_i}\right] \\
    r_{i,sp}(i\xi_n, k_\parallel, \phi) = \frac{1}{\tilde{\Delta}_{i,\delta}}&&\left[\frac{\xi_n K_n}{c}\tilde{\sigma}_{t,i}\delta_i \sin{2\phi_i}\right]\nonumber
\end{eqnarray}
with 
\begin{equation}\label{eq:RcoeffSmall2}
    \tilde{\Delta}_{i,\delta}(i\xi_n, k_\parallel, \phi) = \frac{\xi_n^2}{c^2}\tilde{\sigma}_{t,i} + \frac{\xi_n K_n}{4c}\tilde{\sigma}_{t,i}^2 + K_n^2\tilde{\sigma}_{t,i} + \frac{4\xi_n K_n}{c}
\end{equation}
Let $r_{i, ss/pp} = g_{i, ss/pp} + h_{i, ss/pp}\cdot\delta_i\cos{2\phi_i}$, $r_{i, sp/ps} = h_{i, sp/ps}\cdot\delta_i\sin{2\phi_i}$, then $\mathcal{D}^{(0)}, \mathcal{D}^{(1)}$ and $\mathcal{D}^{(2)}$ are 
\begin{eqnarray}\label{eq:Ds}
    \mathcal{D}^{(0)} && = 1 - (g_{1,ss}g_{2,ss}+g_{1,pp}g_{2,pp})e^{-2K_nd} + g_{1,ss}g_{1,pp}g_{2,ss}g_{2,pp}e^{-4K_nd} \nonumber \\
    \mathcal{D}^{(1)} && = \left[-(h_{1,ss}g_{2,ss} + h_{1,pp}g_{2,pp})e^{-2K_nd} + g_{2,ss}g_{2,pp}(g_{1,ss}h_{1,pp} + h_{1,ss}g_{1,pp})e^{-4K_nd}\right]\cdot\delta_1\cos{2\phi_1}  + [1\leftrightarrow2]\nonumber \\
    \mathcal{D}^{(2)} && = -(h_{1,ss}h_{2,ss} + h_{1,pp}h_{2,pp})e^{-2K_nd}\cdot\delta_1\delta_2\cos{2\phi_1}\cos{2\phi_2} \nonumber \\
    && -(h_{1,sp}h_{2,ps} + h_{1,ps}h_{2,sp})e^{-2K_nd}\cdot\delta_1\delta_2\sin{2\phi_1}\sin{2\phi_2} \nonumber \\
    && + (h_{1,ss}g_{1,pp} + g_{1,ss}h_{1,pp})(h_{2,ss}g_{2,pp} + g_{2,ss}h_{2,pp})e^{-4K_nd}\cdot\delta_1\delta_2\cos{2\phi_1}\cos{2\phi_2} \nonumber \\
    && + h_{1,ss}h_{1,pp}g_{2,ss}g_{2,pp}e^{-4K_nd}\cdot(\delta_1\cos{2\phi_1})^2 + g_{1,ss}g_{1,pp}h_{2,ss}h_{2,pp}e^{-4K_nd}\cdot(\delta_2\cos{2\phi_2})^2 \nonumber \\
    && - h_{1,sp}h_{1,ps}g_{2,ss}g_{2,pp}e^{-4K_nd}\cdot(\delta_1\sin{2\phi_1})^2 - g_{1,ss}g_{1,pp}h_{2,sp}h_{2,ps}e^{-4K_nd}\cdot(\delta_2\sin{2\phi_2})^2
\end{eqnarray}
We mention that only the first three terms in $\mathcal{D}^{(2)}$ will contribute to the torque. Taking Eq.(\ref{eq:Ds}) into Eq.(\ref{eq:CasimirEnergy_2ndOrder}) and integrate over $\phi$, we obtain the Casimir energy
\begin{eqnarray}
    E_c(\theta,d) = && E_{0}(d) + \frac{k_BTA}{4\pi}\sum^{\infty}_{n=0}{}' \int_0^\infty k_\parallel dk_\parallel \frac{1}{(\mathcal{D}^{(0)})^2} \left\{-(h_{1,ss}h_{2,ss} + h_{1,pp}h_{2,pp} + h_{1,sp}h_{2,ps} + h_{1,ps}h_{2,sp})e^{-2K_nd} \right. \nonumber\\
    &&\left. + \left[(g_{1,ss}g_{2,ss}+g_{1,pp}g_{2,pp})(h_{1,sp}h_{2,ps} + h_{1,ps}h_{2,sp}) + 2g_{1,ss}g_{2,ss}h_{1,pp}h_{2,pp} + 2g_{1,pp}g_{2,pp}h_{1,ss}h_{2,ss} \right]e^{-4K_nd} \nonumber\right.\\
    &&\left. - \left[g_{1,ss}g_{1,pp}g_{2,ss}g_{2,pp}(h_{1,sp}h_{2,ps} + h_{1,ps}h_{2,sp}) + g_{1,pp}^2g_{2,pp}^2h_{1,ss}h_{2,ss} + g_{1,ss}^2g_{2,ss}^2h_{1,pp}h_{2,pp}\right]e^{-6K_nd} \right\}\cdot\delta_1\delta_2\cos{2\theta} \nonumber
\end{eqnarray}
where $E_{0}(d)$ is independent of the relative angle between two altermagnets. Taking the derivative with respect to $\theta$, we obtain the Casimir torque
\begin{eqnarray}
    \tau(\theta, d) = && \frac{k_BTA}{2\pi}\sum^{\infty}_{n=0}{}' \int_0^\infty k_\parallel dk_\parallel \frac{1}{(\mathcal{D}^{(0)})^2}
    \left\{-(h_{1,ss}h_{2,ss} + h_{1,pp}h_{2,pp} + h_{1,sp}h_{2,ps} + h_{1,ps}h_{2,sp})e^{-2K_nd} \right.\\
    &&\left. + \left[(g_{1,ss}g_{2,ss}+g_{1,pp}g_{2,pp})(h_{1,sp}h_{2,ps} + h_{1,ps}h_{2,sp}) + 2g_{1,ss}g_{2,ss}h_{1,pp}h_{2,pp} + 2g_{1,pp}g_{2,pp}h_{1,ss}h_{2,ss} \right]e^{-4K_nd} \nonumber\right.\\
    &&\left. - \left[g_{1,ss}g_{1,pp}g_{2,ss}g_{2,pp}(h_{1,sp}h_{2,ps} + h_{1,ps}h_{2,sp}) + g_{1,pp}^2g_{2,pp}^2h_{1,ss}h_{2,ss} + g_{1,ss}^2g_{2,ss}^2h_{1,pp}h_{2,pp}\right]e^{-6K_nd} \right\}\cdot\delta_1\delta_2\sin{2\theta} \nonumber
\end{eqnarray}
The torque is proportional to $\delta_1\delta_2$ at each Matsubara frequency and depends on $\theta$ sinusoidally. 

\subsection{C. Non-retarded limit}
In the non-retarded limit, i.e. $\hbar\frac{c}{d}\gg\hbar\omega_0$, the torque is mainly contributed by the TM wave. Keeping the contribution from $r_{pp}$ only, the Casimir torque in non-retarded and weak anisotropy limit is given by
\begin{eqnarray}
    \tau(\theta, d) = - \frac{k_BTA}{2\pi}\sum^{\infty}_{n=0}{}' \int_0^\infty k_\parallel dk_\parallel \frac{h_{1,pp}h_{2,pp}e^{-2K_nd}}{(1-g_{1,pp}g_{2,pp}e^{-2K_nd})^2}\delta_1\delta_2\sin{2\theta}
\end{eqnarray}
with
\begin{eqnarray}
    && g_{i,pp} = \frac{1}{\tilde{\Delta}_{i,\delta}}\left[\frac{\xi_n K_n}{4c}\tilde{\sigma}_{t,i}^2 + K_n^2\tilde{\sigma}_{t,i}\right]\\
    && h_{i,pp} = \frac{K_n^2\tilde{\sigma}_{t,i}}{\tilde{\Delta}_{i,\delta}} - \frac{1}{\tilde{\Delta}^2_{i,\delta}}\left[\frac{\xi_n K_n}{4c}\tilde{\sigma}_{t,i}^2 + K_n^2\tilde{\sigma}_{t,i}\right]\left[-\frac{\xi_n^2}{c^2}\tilde{\sigma}_{t,i} + K_n^2\tilde{\sigma}_{t,i}\right]\\
    && \tilde{\Delta}_{i,\delta} = \frac{\xi_n^2}{c^2}\tilde{\sigma}_{t,i} + \frac{\xi_n K_n}{4c}\tilde{\sigma}_{t,i}^2 + K_n^2\tilde{\sigma}_{t,i} + \frac{4\xi_n K_n}{c}
\end{eqnarray}
In our case, $\tilde{\sigma}_{t,i}$ is also a small quantity. Changing the integration variable to $\tilde{K} = K_nd$ and using the property that $\frac{\xi_nd}{c}\ll1, \tilde{\sigma}_{t,i}\ll1$, the torque formula is written as
\begin{eqnarray}
    \tau(\theta, d) = - \frac{k_BTA}{2\pi d^2}\sum^{\infty}_{n=0}{}' \int_0^\infty \tilde{K}d\tilde{K} \frac{h_{1,pp}h_{2,pp}e^{-2\tilde{K}}}{(1-g_{1,pp}g_{2,pp}e^{-2\tilde{K}})^2}\delta_1\delta_2\sin{2\theta}
\end{eqnarray}
with
\begin{eqnarray}
    && g_{i,pp} = \frac{\tilde{K}\tilde{\sigma}_{t,i}}{\tilde{K}\tilde{\sigma}_{t,i} + \frac{4\xi_nd}{c}}\\
    && h_{i,pp} = \frac{\tilde{K}\tilde{\sigma}_{t,i}\cdot\frac{4\xi_nd}{c}}{(\tilde{K}\tilde{\sigma}_{t,i} + \frac{4\xi_nd}{c})^2}
\end{eqnarray}
Since $\frac{\xi_nd}{c}, \tilde{\sigma}_{t,i}$ are both small quantities, we can not neglect either term in the denominators of $g_{i,pp}, h_{i,pp}$. Thus, the torque formula can not be expressed as a Laurent polynomial function of $\xi_nd/c$ as in the case of uniaxial materials \cite{2017Munday}, which makes the scaling ambiguous.

\subsection{D. High-temperature limit}
In high temperature limit, i.e. $\omega_0 \gg k_BT\gg\hbar\frac{c}{d}$, the dominant contribution to the torque comes from the $n=1$ Matsubara term. Different from the Casimir torque between bulk anisotropic materials, the reflection coefficients for two-dimensional altermagnets at the lowest Matsubara frequency term $n = 0$ are $r_{ss}=r_{sp}=r_{ps}=0$, $r_{pp}=1$ and have no angular dependence. As a result, the $n = 0$ term makes a vanishing contribution to the torque. Considering only $n=1$ Matsubara term and using one-reflection approximation, the torque is given by
\begin{eqnarray}
\tau_c(\theta, d) = -\frac{k_BTA}{32\pi} \left[\int_0^\infty k_\parallel dk_\parallel (\frac{\xi_1^2}{c^2K_1^2} + \frac{c^2K_1^2}{\xi_1^2} - 2) e^{-2K_1d}\right]
\cdot(\sigma_{xx,1} - \sigma_{yy,1})(\sigma_{xx,2} - \sigma_{yy,2})\cdot\sin{2\theta}
\end{eqnarray}
where $K_1 = \sqrt{\xi_1^2/c^2 + k^2_\parallel}$, and $\sigma_{ij}=\sigma_{ij}(i\xi_1)$. Replacing the integration variable $k_\parallel$ by $u = 2K_1d$, the integration part can be written as
\begin{eqnarray}
&& \int_0^\infty k_\parallel dk_\parallel (\frac{\xi_1^2}{c^2K_1^2} + \frac{c^2K_1^2}{\xi_1^2} - 2) e^{-2K_1d} \\
  = && \frac{\xi_1^2}{c^2}\int_{2\frac{\xi_1}{c}d}^\infty du\frac{e^{-u}}{u} 
  + \frac{1}{(2d)^4}\frac{c^2}{\xi_1^2}\int_{2\frac{\xi_1}{c}d}^\infty du u^3e^{-u} 
  - \frac{2}{(2d)^2} \int_{2\frac{\xi_1}{c}d}^\infty du ue^{-u}\nonumber\\
  = && \left\{\frac{\xi_1^2}{c^2} \textup{E}_1(2\frac{\xi_1}{c}d) 
  + \frac{1}{(2d)^4}\frac{c^2}{\xi_1^2}\left[(2\frac{\xi_1}{c}d)^3 + 3(2\frac{\xi_1}{c}d)^2 + 6(2\frac{\xi_1}{c}d) + 6\right]
  - \frac{2}{(2d)^2}(2\frac{\xi_1}{c}d+1)\right\}e^{-2\frac{\xi_1}{c}d}\nonumber\\
  \simeq && \left\{\frac{\xi_1^2}{c^2}\left[\frac{1}{(2\frac{\xi_1}{c}d)} - \frac{1}{(2\frac{\xi_1}{c}d)^2} + \frac{2}{(2\frac{\xi_1}{c}d)^3}\right]
  + \frac{1}{(2d)^4}\frac{c^2}{\xi_1^2}\left[(2\frac{\xi_1}{c}d)^3 + 3(2\frac{\xi_1}{c}d)^2 + 6(2\frac{\xi_1}{c}d) + 6\right]
  - \frac{2}{(2d)^2}(2\frac{\xi_1}{c}d+1)\right\}e^{-2\frac{\xi_1}{c}d}\nonumber\\
  \simeq && \frac{c}{\xi_1} \frac{e^{-2\frac{\xi_1}{c}d}}{d^3} \nonumber
\end{eqnarray}
where $\textup{E}_1(x)=\int_x^\infty \frac{e^{-t}}{t}dt$. To obtain the fourth line, we have used the asymptotic property of $\textup{E}_1(x)$ that $\textup{E}_1(x)\approx\frac{e^{-x}}{x}(1-\frac{1}{x}+\frac{2}{x^2})$ when $x\gg1$. Therefore, the torque is given by
\begin{eqnarray}
\tau_c(\theta,d) \simeq -\frac{\hbar c A}{64 \pi^2} \frac{e^{-2\frac{\xi_1}{c}d}}{d^3}
(\sigma_{xx,1} - \sigma_{yy,1})(\sigma_{xx,2} - \sigma_{yy,2})\sin{2\theta}
\end{eqnarray}
The torque scales as a function of separation $d$ as $\tau_c\sim {e^{-2\frac{\xi_1}{c}d}}/{d^3}$ in the retarded and high temperature limit.


\subsection{E. Retarded limit}
In the retarded and low temperature limit, i.e. $\hbar\omega_0\gg\hbar\frac{c}{d}\gg k_BT$, the Matsubara summation can be replaced by an integration over the continuous variable $\xi$. Defining the dimensionless variables $\tilde{\xi} = \xi d/c$ and $\tilde{k} = kd$, Eq.(\ref{eq:TorqueGeneral}) becomes
\begin{eqnarray}
    \tau(\theta, d) = \frac{\hbar c A}{8\pi^3d^3}\int^{\infty}_{0}d\tilde{\xi} \int_0^\infty \tilde{k}_\parallel d\tilde{k}_\parallel \int_0^{2\pi} d\phi \; 
    \textup{Tr}\left[(1-\textbf{R}_1 \cdot \textbf{R}_2 e^{-2\tilde{K}_n})^{-1}\textbf{R}_1 \partial_\theta\textbf{R}_2\right]e^{-2\tilde{K}_n}
\end{eqnarray}
where $\tilde{K}_n = \sqrt{\tilde{\xi}^2 + \tilde{k}_\parallel^2}$ and $\textbf{R}_i(i\frac{\tilde{\xi}}{d}c, \frac{\tilde{k}_\parallel}{d}, \phi_i)$. In the retarded limit, the conductivity satisfies $\sigma(i\frac{\xi}{d}c)\approx\sigma(0)$, so the reflection matrix $\textbf{R}_i(i\frac{\tilde{\xi}}{d}c, \frac{\tilde{k}_\parallel}{d}, \phi_i) = \textbf{R}_i(i\tilde{\xi}c, \tilde{k}_\parallel, \phi_i)$. 
Because both $\tilde{K}_n$ and $\mathbf{R}_i$ are independent of the separation $d$, the integrand shows no distance dependence, leading to the characteristic $\tau \sim 1/d^3$ scaling of the torque. 

For our case, the cutoff response of the altermagnet $\hbar\omega_0\sim 1\textup{eV}$. At temperature $T = 30K$, the thermal energy $k_BT \approx 2.6\textup{meV}$, thus the distance range satisfying $0.2\mu m \ll d \ll 80 \mu m$ belongs to the retarded and low temperature limit, where the torque exhibits a scaling $\tau \sim 1/d^3$. At higher temperature, e.g. at $T = 300K$, no distance regime satisfies $\hbar\omega_0\gg\hbar\frac{c}{d}\gg k_BT$ (or equivalently, $0.2\mu m \ll d \ll 8 \mu m$), thus eliminating the pure $1/d^3$ retarded regime. Instead, the torque decays faster than $1/d^3$ at intermediate distances and ultimately transitions to the high temperature scaling $\exp(-2\xi_1 d / c)/d^3$ at large distances.

\section{III. \label{Appendix:Rcoeff}The reflection coefficients for the altermagnet thin film}
In this section, we derive the reflection coefficients of the electromagnetic waves incident on the vacuum-altermagnet interface. We first consider the case where the altermagnetic thin film is placed on an isotropic substrate with dielectric function $\epsilon$ and magnetic susceptibility $\mu$. Taking the limit $\epsilon\rightarrow\epsilon_0, \mu\rightarrow\mu_0$, we obtain the reflection coefficients of the altermagnet thin film standing in vacuum. We then introduce the degree of anisotropy $\delta = \frac{\sigma_{xx} - \sigma_{yy}}{\sigma_{xx} + \sigma_{yy}}$ and the total diagonal conductivity $\sigma_t = \sigma_{xx} + \sigma_{yy}$, and express the reflection coefficients for 2D altermagnets in terms of these quantities. 

When the wave vector of incident electromagnetic waves $\mathbf{k} = (k_x, 0, k_z)$ lies in the x-O-z plane, the incident electromagnetic waves can be written as:
\begin{eqnarray}
    \mathbf{E}_{in} = \left[ A_\perp\hat{y}+\frac{cA_\parallel}{\omega}(k_z\hat{x}-k_x\hat{z}) \right]e^{ik_xx+ik_zz-i\omega t} \\
    \mathbf{H}_{in} = \left[ \frac{A_\perp}{\mu_0\omega}(k_x\hat{z}-k_z\hat{x}) + \epsilon_0 c A_\parallel\hat{y} \right]e^{ik_xx+ik_zz-i\omega t}
\end{eqnarray}
Here, $A_\perp (A_\parallel)$ is the magnitude of the incident TE (TM) wave and has the dimension of the electric field.
The reflected electromagnetic waves can be written as:
\begin{eqnarray}
    \mathbf{E}_{r} = \left[ R_\perp\hat{y}-\frac{c R_\parallel}{\omega}(k_z\hat{x}+k_x\hat{z}) \right]e^{ik_xx-ik_zz-i\omega t} \\
    \mathbf{H}_{r} = \left[ \frac{R_\perp}{\mu_0\omega}(k_x\hat{z}+k_z\hat{x}) + \epsilon_0 c R_\parallel\hat{y} \right]e^{ik_xx-ik_zz-i\omega t}
\end{eqnarray}
where $R_\perp (R_\parallel)$ is the magnitude of the reflected TE (TM) wave. Next, we find the two modes of the electromagnetic waves in the substrate. Suppose the electromagnetic waves can be written as:
\begin{eqnarray}
    \mathbf{E}_{sub} & = (e_x\hat{x} + e_y\hat{y} + e_z\hat{z})e^{ik_xx+iqz-i\omega t} \\
    \mathbf{H}_{sub} & = (h_x\hat{x} + h_y\hat{y} + h_z\hat{z})e^{ik_xx+iqz-i\omega t}
\end{eqnarray}
Putting $\mathbf{E}_{sub}, \mathbf{H}_{sub}$ into the Maxwell's equations, we can get the two electromagnetic wave modes: $\mathbf{e}_1 = (e_x, 0, -\frac{k_x}{q}e_x)$, $\mathbf{h}_1 = (0, \frac{\omega}{q}\epsilon e_x, 0)$, and $\mathbf{e}_2 = (0, e_y, 0), \mathbf{h}_2 = (-\frac{q}{\mu\omega}e_y, 0, \frac{k_x}{\mu\omega}e_y)$ with $q =  \epsilon\mu\omega^2 - k_x^2$. In the end, the electromagnetic waves in the substrate can be written as:
\begin{eqnarray}
    \mathbf{E}_{sub} = \left(e_x\hat{x} + e_y\hat{y} -\frac{k_x}{q}e_x \hat{z} \right)e^{ik_xx+iqz-i\omega t} \\
    \mathbf{H}_{sub} = \left(-\frac{q}{\mu\omega}e_y\hat{x} + \frac{\omega}{q}\epsilon e_x\hat{y} + \frac{k_x}{\mu\omega}e_y\hat{z}   \right)e^{ik_xx+iqz-i\omega t}
\end{eqnarray}

Next, we derive the reflection coefficients. The boundary conditions are $\hat{\mathbf{n}}\times(\mathbf{E}_2 - \mathbf{E}_1) = 0, \hat{\mathbf{n}}\times(\mathbf{H}_2 - \mathbf{H}_1) = \mathbf{J_s}$, where $\mathbf{J_s} = \mathbf{\sigma_s}\mathbf{E}$ is the surface current and $\mathbf{\sigma_s}$ is the conductivity of the altermagnetic thin film. $\hat{\mathbf{n}}$ is the unit vector pointing in the normal direction from substrate 1 to substrate 2. Written out explicitly, the boundary conditions are
\begin{eqnarray}
    (A_\parallel - R_\parallel)\frac{c}{\omega}k_z &=& e_x \\
    A_\perp + R_\perp &=& e_y \\
    (-A_\perp + R_\perp)\frac{k_z}{\mu_0\omega} &=& h_x - \sigma_{yx} e_x - \sigma_{yy} e_y \\
    (A_\parallel + R_\parallel)\epsilon_0 c &=& h_y + \sigma_{xy} e_y + \sigma_{xx} e_x
\end{eqnarray}
where $h_x = -\frac{q}{\mu\omega}e_y, h_y = \frac{\omega}{q}\epsilon e_x$. Solving the equations, we obtain the reflection coefficients
\begin{eqnarray}
    r_{ss} &=& \frac{-\mu\omega[\omega\sigma_{yy}\mu_0(\epsilon_0 q + \epsilon k_z) + \mu_0qk_z(\sigma_{xx}\sigma_{yy} - \sigma_{xy}\sigma_{yx})] + (\mu k_z - \mu_0 q)[\epsilon_0q\omega + k_z(q\sigma_{xx} + \omega\epsilon)]}{\Delta} \\  
    r_{pp} &=& \frac{\mu\omega[\omega\sigma_{yy}\mu_0(-\epsilon_0 q + \epsilon k_z) + \mu_0qk_z(\sigma_{xx}\sigma_{yy} - \sigma_{xy}\sigma_{yx})] + (\mu k_z + \mu_0q)[-\epsilon_0 q\omega + k_z(q\sigma_{xx} + \omega\epsilon)]}{\Delta} \\
    r_{ps} &=& \frac{2\sigma_{xy}qk_z\mu\omega / c}{\Delta} \\ 
    r_{sp} &=& -\frac{2\sigma_{yx}qk_z\mu\omega / c}{\Delta}
\end{eqnarray}
where $\Delta = \mu\omega[\omega\sigma_{yy}\mu_0(\epsilon_0 q + \epsilon k_z) + \mu_0qk_z(\sigma_{xx}\sigma_{yy} - \sigma_{xy}\sigma_{yx})] + (\mu k_z + \mu_0q)[\epsilon_0q\omega + k_z(q\sigma_{xx} + \omega\epsilon)]$. 
When the altermagnetic thin film is standing in vacuum, i.e. $\epsilon\rightarrow\epsilon_0, \mu\rightarrow\mu_0$, the reflection coefficients become
\begin{eqnarray}\label{Rcoeff}
    r_{ss} &=& \frac{-\mu_0\omega[2\omega\sigma_{yy}\epsilon_0 + k_z(\sigma_{xx}\sigma_{yy} - \sigma_{xy}\sigma_{yx})]}{\Delta} \nonumber\\  
    r_{pp} &=& \frac{\mu_0\omega k_z(\sigma_{xx}\sigma_{yy} - \sigma_{xy}\sigma_{yx}) + 2k_z^2\sigma_{xx}}{\Delta} \nonumber\\
    r_{ps} &=& \frac{2\sigma_{xy}k_z\omega / c}{\Delta} \\ 
    r_{sp} &=& -\frac{2\sigma_{yx}k_z\omega / c}{\Delta} \nonumber
\end{eqnarray}
where $\Delta = \mu_0\omega[2\omega\sigma_{yy}\epsilon_0 + k_z(\sigma_{xx}\sigma_{yy} - \sigma_{xy}\sigma_{yx})] + 2k_z(2\epsilon_0\omega + k_z\sigma_{xx})$. 

When the electromagnetic waves incident from other directions with $\mathbf{k} = (k_x, k_y, k_z)$, the reflection coefficients should be modified by changing $\sigma_{ij}$ into $\sigma_{ij}(\phi)$, where $\phi$ is the angle between the in-plane wave vector $\mathbf{k}_\parallel = (k_x, k_y, 0)$ and the x-axis. $\sigma_{ij}(\phi)$ is the conductivity tensor of the altermagnet in the new coordinate system rotated counter-clockwise by angle $\phi$, which is given by
\begin{equation}
    \sigma(\phi) = \left(
    \begin{array}{ccc}
    \sigma_{xx}\tilde{c}^2 + \sigma_{yy}s^2 + (\sigma_{xy} + \sigma_{yx})s\tilde{c} & (-\sigma_{xx} + \sigma_{yy})s\tilde{c} + \sigma_{xy}\tilde{c}^2 - \sigma_{yx}s^2 \\
    (-\sigma_{xx} + \sigma_{yy})s\tilde{c} - \sigma_{xy}s^2 + \sigma_{yx}\tilde{c}^2 & \sigma_{xx}s^2 + \sigma_{yy}\tilde{c}^2 - (\sigma_{xy} + \sigma_{yx})s\tilde{c}
    \end{array}
    \right)
\end{equation}
where $\sigma_{ij}$ is the conductivity at $\phi = 0$, $s = \sin\phi, \tilde{c} = \cos\phi$. 

In our case, the symmetry of the altermagnet guarantees that the off-diagonal conductivity $\sigma_{xy}, \sigma_{xy}$ is zero. Therefore, 
\begin{equation}
    \sigma(\phi) = \left(
    \begin{array}{ccc}
    \sigma_{xx}\tilde{c}^2 + \sigma_{yy}s^2 & (\sigma_{yy}-\sigma_{xx})s\tilde{c}\\
    (\sigma_{yy}-\sigma_{xx})s\tilde{c} & \sigma_{xx}s^2 + \sigma_{yy}\tilde{c}^2
    \end{array}
    \right)
\end{equation}
Since $\sigma_{xx}(\phi)\sigma_{yy}(\phi) - \sigma_{xy}(\phi)\sigma_{yx}(\phi) = \sigma_{xx}\sigma_{yy}$ is independent of $\phi$, the reflection coefficients of electromagnetic waves in an arbitrary wave vector can be written as 
\begin{eqnarray}
    r_{ss}(\omega, k_\parallel, \phi) &=& \frac{-\mu_0\omega[2\omega\epsilon_0(\sigma_{xx}s^2 + \sigma_{yy}\tilde{c}^2) + k_z\sigma_{xx}\sigma_{yy}]}{\Delta} \nonumber\\  
    r_{pp}(\omega, k_\parallel, \phi) &=& \frac{\mu_0\omega k_z\sigma_{xx}\sigma_{yy} + 2k_z^2(\sigma_{xx}\tilde{c}^2 + \sigma_{yy}s^2)}{\Delta} \nonumber\\
    r_{ps}(\omega, k_\parallel, \phi) &=& \frac{2 (\sigma_{yy}-\sigma_{xx})s\tilde{c}k_z\omega / c}{\Delta} \\ 
    r_{sp}(\omega, k_\parallel, \phi) &=& -\frac{2 (\sigma_{yy}-\sigma_{xx})s\tilde{c}k_z\omega / c}{\Delta} \nonumber
\end{eqnarray}
where $\Delta = \mu_0\omega[2\omega\epsilon_0(\sigma_{xx}s^2 + \sigma_{yy}\tilde{c}^2) + k_z\sigma_{xx}\sigma_{yy}] + 4k_z\omega\epsilon_0 + 2k_z^2(\sigma_{xx}\tilde{c}^2 + \sigma_{yy}s^2)$.

Defining $\delta = \frac{\sigma_{xx} - \sigma_{yy}}{\sigma_{xx} + \sigma_{yy}}$ and $\sigma_t = \sigma_{xx} + \sigma_{yy}$, the conductivity $\sigma(\phi)$ can be written as
\begin{equation}
    \sigma(\phi) = \left(
    \begin{array}{ccc}
     \frac{1}{2} + \frac 1 2 \delta\cos{2\phi} & -\frac 1 2 \delta\sin{2\phi}\\
     -\frac 1 2 \delta\sin{2\phi} & \frac{1}{2} - \frac 1 2 \delta\cos{2\phi}
    \end{array} 
    \right) \times \sigma_t
\end{equation}
and the reflection coefficients are given by
\begin{eqnarray}
    r_{ss}(\omega, k_\parallel, \phi) &=& -\frac{1}{\Delta_\delta}\left[\frac{\omega^2}{c^2}\tilde{\sigma}_t(1 - \delta\cos{2\phi}) + \frac{\omega k_z}{4c}\tilde{\sigma}_t^2(1-\delta^2)\right]\nonumber\\  
    r_{pp}(\omega, k_\parallel, \phi) &=& \frac{1}{\Delta_\delta}\left[\frac{\omega k_z}{4c}\tilde{\sigma}_t^2(1-\delta^2) + k_z^2\tilde{\sigma}_t(1 + \delta\cos{2\phi})\right]\nonumber\\
    r_{ps}(\omega, k_\parallel, \phi) &=& -\frac{1}{\Delta_\delta}\left[\frac{\omega k_z}{c}\tilde{\sigma}_t\delta \sin{2\phi}\right] \\
    r_{sp}(\omega, k_\parallel, \phi) &=& \frac{1}{\Delta_\delta}\left[\frac{\omega k_z}{c}\tilde{\sigma}_t\delta \sin{2\phi}\right]\nonumber
\end{eqnarray}
where $\Delta_\delta = \frac{\omega^2}{c^2}\tilde{\sigma}_t(1 - \delta\cos{2\phi}) + \frac{\omega k_z}{4c}\tilde{\sigma}_t^2(1-\delta^2) + k_z^2\tilde{\sigma}_t(1 + \delta\cos{2\phi}) + \frac{4\omega k_z}{c}$. Here, the normalized total conductivity $\tilde{\sigma}_t = \frac{\sigma_t}{\sigma_0}$ and $\sigma_0 = \sqrt{\frac{\epsilon_0}{\mu_0}}$. 

\bibliography{ref} 